\documentclass[12pt]{article}

\usepackage{amssymb}
\usepackage{amsmath}

\newcommand{\beq}{\begin{equation}}
\newcommand{\eeq}{\end{equation}}
\newcommand{\bea}{\begin{eqnarray}}
\newcommand{\eea}{\end{eqnarray}}

\newcommand{\e}{\mbox{e}}
\renewcommand{\d}{\mbox{d}}

\newcommand{\La}{\Lambda}

\renewcommand{\th}{\theta}
\newcommand{\Del}{\Delta}

\newcommand{\ra}{\rangle}
\newcommand{\la}{\langle}

\newcommand{\mi}{\!-\!}
\newcommand{\equ}{\!=\!}

\newcommand{\cD}{{\cal D}}

\newcommand{\bX}{{\bar{X}}}

\newcommand{\SL}{\sqrt{\La}}

\newcommand{\eref}[1]{\eqref{#1}}

\begin{document}

\begin{center}
{\normalsize \hfill SPIN-06/36}\\
{\normalsize \hfill ITP-UU-06/46}\\
{\normalsize \hfill IMPERIAL-TP-06-SZ-04}\\

\vspace{60pt}
{ \Large \bf The emergence of $AdS_2$ 
from quantum fluctuations}

\vspace{30pt}

{\sl J. Ambj\o rn}$\,^{a,c}$
{\sl R. Janik}$\,^{b}$,
{\sl W. Westra}$\,^{c}$
and {\sl S. Zohren}$\,^{d}$

\vspace{30pt}
{\footnotesize

$^a$~The Niels Bohr Institute, Copenhagen University\\
Blegdamsvej 17, DK-2100 Copenhagen \O , Denmark.\\
{ email: ambjorn@nbi.dk}\\

\vspace{10pt}

$^b$~
Institute of Physics, Jagellonian University,\\
ul. Reymonta 4, 30-059 Krakow, Poland.\\
email: ufrjanik@if.uj.edu.pl\\

\vspace{10pt}

$^c$~Institute for Theoretical Physics, Utrecht University, \\
Leuvenlaan 4, NL-3584 CE Utrecht, The Netherlands.\\
email: w.westra@phys.uu.nl, ambjorn@phys.uu.nl\\

\vspace{10pt}

$^d$~Blackett Laboratory, Imperial College,\\
London SW7 2AZ, United Kingdom.\\
email: stefan.zohren@imperial.ac.uk\\
}
\vspace{30pt}

\begin{abstract}
We have shown how the quantization of two-dimensional quantum gravity with an action which contains only a positive cosmological constant and boundary cosmological constants leads to the emergence of a spacetime which can be described as a constant negative curvature spacetime with superimposed quantum fluctuations.
\end{abstract}

\vspace{40pt}

{\footnotesize Talk given by W. Westra at the Eleventh Marcel Grossmann Meeting on General Relativity at the Freie U. Berlin, July 23 - 29, 2006.}

\end{center}
\newpage

\section{Introduction}\label{intro}

The causal dynamical triangulations approach to quantum gravity (CDT) is an attempt to define the gravitational path integral in a background independent and nonperturbative manner. As in the case of Euclidean dynamical triangulation approaches (DT), CDT provides a regularization of the path integral through a sum over piecewise linear geometries where the edge length of the individual building blocks serves as an ultraviolet cutoff. However, in contrast to DT a global time foliation is imposed on each individual history in the path integral. 

The method of CDT was first applied to two-dimensional quantum gravity where the model was shown to be analytically solvable \cite{Ambjorn:1998xu}.
Although two-dimensional quantum gravity does not have any propagating degrees of freedom it is a fertile playground to study certain aspects of diffeomorphism invariant theories. Among the issues that have been addressed within the two-dimensional framework are the inclusion of a sum over topologies \cite{Loll:2005dr} and the emergence of a background geometry purely from quantum fluctuations \cite{Ambjorn:2006hu} where the latter is the subject of this article.

One of the most natural quantities to study in CDT is the loop-loop-propagator which is the amplitude for a transition from a spacelike loop with boundary cosmological constant $X$ to a loop with boundary cosmological constant $Y$ in time $T$,   
\beq\label{2.a0}
G_\La (X,Y;T) =
\int \cD [g] \; e^{-S[g]},\,\,\, S[g] = \La \int \d^2x \sqrt{g(x,t)} +
X \oint \d l_1 + Y \oint \d l_2,
\eeq
where the action only includes bulk and boundary cosmological constants, since the curvature term in the Einstein-Hilbert action is trivial in 2D.

Evaluating the path integral using the CDT regularization and taking the continuum limit yields the following \cite{Ambjorn:1998xu}
\beq\label{2.a3}
G_\La (X,Y;T) = \frac{\bX^2(T,X)-\La}{X^2-\La} \; \frac{1}{\bX(T,X)+Y},
\eeq
where $\bX(T,X)$ is the solution of
\beq\label{2.3}
\frac{\d \bX}{\d T} = -(\bX^2-\La),~~~\bX(0,X)=X.
\eeq

\section{The emergence of $AdS_2$}

To determine the background geometry of the 1+1 dimensional universe we calculate the average spatial length at proper time $t\in[0,T]$
\beq\label{2.a7}
\la L(t)\ra_{X,Y,T} =
\frac{1}{G_\La (X,Y;T)} \int_0^\infty \d L\;
G_\La (X,L;t) \;L\; G_\La (L,Y;T-t).
\eeq
Evaluating the average length at the boundary $t = T$ and taking the limit $T\to\infty$ gives 
\beq\label{2.a8}
\lim_{T\to\infty}\la L(T)\ra_{X,Y,T} =\frac{1}{Y+\SL}.
\eeq
Interestingly, one observes that there is a special value
$Y \equ \mi \SL$ of the boundary cosmological constant for which the boundary length diverges and the geometry becomes non-compact. Using this critical value for the boundary cosmological constant $Y$ one can obtain the boundary length for finite $T$
\beq\label{5.2}
L^c(T) = 
\la L(T) \ra_{X,Y= -\SL;T} = \frac{1}{\SL} \; \frac{1}{\coth \SL T -1}.
\eeq
Instead of using boundary cosmological constants one can also fix the spatial length of the boundaries.
Using the Laplace transformed propagator $G_\La(L_1,L_2;T)$ we can evaluate the average spatial length $\la L(t)\ra_{L_1,L_2,T}$ for fixed lengths $L_1$ and $L_2$ of the boundary loops. 

In the following we want to investigate the quantum geometry in the case where it becomes non-compact. Therefore we set the boundary length at $t=T$ to the critical value $L^c(T)$ as defined in \eref{5.2} and for simplicity we shrink the spatial geometry at $t=0$ to a point. In the limit $T\to\infty$ one obtains the average length of the spatial geometry at proper time $t\in[0,T]$
\beq\label{2.a9}
\la L(t) \ra \equiv \lim_{T\to\infty} \la L(t)\ra_{L_1=0,L_2=L^c(T),T}  = \frac{1}{\SL} \; \sinh (2\SL t).
\eeq
Due to the fact that $L$ and $T$ are defined from the continuum limit of a simplicial geometry there is a relative constant of proportionality that can only be fixed by comparing with continuum calculations \cite{Nakayama:1993we} yielding $L_{cont}(t)=\pi\la L(t) \ra$.
From this result the metric for the background geometry is readily obtained,
\beq\label{2.a11}
\d s^2 = \d t^2 + \frac{L_{cont}^2}{4\pi^2}\; \d \th^2 =
\d t^2 + \frac{\sinh^2 (2\SL t)}{4 \La} \;\d \th^2.
\eeq
This is nothing but the metric of the Poincar\'e disc which can be seen as a Wick rotated version of $AdS_2$ with constant negative curvature $R=-8\La$.

To better understand the quantum nature of the geometry it is useful to compute the fluctuations of the spatial length. From expressions analogous to \eref{2.a7} one can determine the relative fluctuations 
\beq\label{5.a5}
\frac{\Del L(t)}{\la L(t)\ra} \equiv  \frac{\sqrt{\la L^2(t)\ra -\la L(t)\ra^2}}{\la L(t)\ra}
\sim \e^{-\SL t}.
\eeq
Surprisingly, the fluctuations of the spatial geometry become exponentially small for $t\gg \La^{-1/2}$. Concluding from Eqs. (\ref{2.a11}) and (\ref{5.a5}), one can view the quantum geometry as a version of Wick rotated $AdS_2$ dressed with small quantum fluctuations.

\section{Discussion}

We have shown that in 2D quantum gravity defined through CDT there is a transition from compact geometry to non-compact $AdS_2$-like geometry for a special value of the boundary cosmological constant. This phenomenon is similar to the Euclidean case where non-compact ZZ-branes appear in a transition from compact 2D geometries in Liouville quantum gravity \cite{Ambjorn:2004my}. A surprising feature of the CDT result is that the fluctuations become exponentially small which enables us to interpret the emerging $AdS_2$ spacetime as a genuine semiclassical background. It is interesting that similar results have been reported in four-dimensional CDT where numerical simulations indicate the emergence of a semi-classical background from a nonperturbative and background-independent path integral \cite{Ambjorn:2005qt}.

\section*{Acknowledgments}
All authors acknowledge support by
ENRAGE (European Network on
Random Geometry), a Marie Curie Research Training Network in the
European Community's Sixth Framework Programme, network contract
MRTN-CT-2004-005616.


\begin{thebibliography}{1}

\bibitem{Ambjorn:1998xu}
J.~Ambj{\o}rn and R.~Loll, {\em Nucl. Phys. B} {\bf 536}, 407 (1998)
[hep-th/9805108].
%%CITATION = HEP-TH 9805108;%%

\bibitem{Loll:2005dr}
R.~Loll, W.~Westra and S.~Zohren, {\em Nucl. Phys. B} {\bf 751}, 419 (2006)
  [hep-th/0507012].
  %%CITATION = HEP-TH 0507012;%%

\bibitem{Ambjorn:2006hu}
J.~Ambj{\o}rn, R.~Janik, W.~Westra and S.~Zohren, {\em Phys. Lett. B} {\bf
  641}, 94 (2006)
    [gr-qc/0607013].
  %%CITATION = GR-QC 0607013;%%

\bibitem{Nakayama:1993we}
R.~Nakayama, {\em Phys. Lett. B} {\bf 325}, 347 (1994)
  [hep-th/9312158].
  %%CITATION = HEP-TH 9312158;%%

\bibitem{Ambjorn:2004my}
J.~Ambj{\o}rn, S.~Arianos, J.~A. Gesser and S.~Kawamoto, {\em Phys. Lett. B} {\bf
  599}, 306 (2004)
  [hep-th/0406108].
 %%CITATION = HEP-TH 0406108;%%

\bibitem{Ambjorn:2005qt}
J.~Ambj{\o}rn, J.~Jurkiewicz and R.~Loll, 
{\em Phys.\ Rev.\ Lett.}  {\bf 93}, 131301 (2004)
  [arXiv:hep-th/0404156],
  %%CITATION = HEP-TH 0404156;%%
{\em Phys.\ Lett.\ B} {\bf 607}, 205 (2005) 
[hep-th/0411152],
%%CITATION = HEP-TH 0411152;%% 
{\em Phys. Rev.} {\bf D72}, p. 064014 (2005) 
[hep-th/0505154].
%%CITATION = HEP-TH 0505154;%%.

\end{thebibliography}
\end{document}